\documentclass{aa}

\usepackage{graphicx}
\usepackage{txfonts}
\usepackage{color}

\def\arcsec{\hbox{$^{\prime\prime}$}}

\bibpunct[,]{(}{)}{;}{a}{}{,}

\begin{document}

\title{The distribution of stars around the Milky Way's black hole}
\subtitle{III. Comparison with simulations}

\author{H. Baumgardt\inst{1}
        \and
        P. Amaro-Seoane\inst{2}
        \and
        R. Sch\"odel\inst{3}} 

 \institute{School of Mathematics and Physics, University of Queensland, St. Lucia, QLD 4072, Australia
  \email{h.baumgardt@uq.edu.au}
 \and
Institute of Space Sciences (ICE, CSIC) \& Institut d'Estudis Espacials de Catalunya (IEEC)\\
at Campus UAB, Carrer de Can Magrans s/n 08193 Barcelona, Spain\\
Institute of Applied Mathematics, Academy of Mathematics and Systems Science, CAS, Beijing 100190, China\\
Kavli Institute for Astronomy and Astrophysics, Beijing 100871, China\\
Zentrum f{\"u}r Astronomie und Astrophysik, TU Berlin, Hardenbergstra{\ss}e 36, 10623 Berlin, Germany
  \email{pau@ice.cat}
 \and
  Instituto de Astrof\'isica de Analuc\'ia, Glorieta de la Astronom\'ia s/n, 18008 Granada, Spain
}

\date{draft \today}

\abstract
{The distribution of stars around a massive black hole (MBH) has been addressed
in stellar dynamics for the last four decades by a number of authors. Because of its proximity,
the centre of the Milky Way is the only observational test case where the stellar distribution can be
accurately tested. Past observational work indicated that the brightest
giants in the Galactic Centre (GC) may show a density deficit around the
central black hole, not a cusp-like distribution, while we
theoretically expect the presence of a stellar cusp.}
{We here present a solution to this long-standing problem.}
{We performed direct-summation $N-$body simulations of star clusters around massive black holes and compared the results
of our simulations with new observational data of the GC's nuclear cluster.}
{We find that after a Hubble time, the distribution of bright stars as well as the diffuse light
follow power-law distributions in projection with slopes of $\Gamma \approx 0.3$
in our simulations. This is in excellent agreement with what is seen in star counts and in the
distribution of the diffuse stellar light extracted from adaptive-optics (AO) assisted near-infrared observations of the GC.}
{Our simulations also confirm that there exists a missing giant star population within a projected radius of a few arcsec around Sgr A*. Such a depletion of giant stars in the innermost 0.1 pc could be explained by a previously present gaseous disc and collisions, which means that a stellar cusp would also be present at the innermost radii, but in the form of degenerate compact cores.}


\maketitle

\section{Introduction}
\label{sec.intro}

The Galactic centre (GC) is the nucleus closest to us, offering the
unique possibility of studying the interaction between a supermassive
black hole (SMBH) and its surrounding star cluster using individual resolved stars. Adaptive-optics (AO) assisted near-infrared photometric and
spectroscopic studies of the stars within about 1\,pc of the SMBH have
shown that the surface density of massive early-type stars is rising
steeply towards Sagittarius\,A* (Sgr\,A*, the electromagnetic
manifestation of the SMBH), while the density distribution of bright late-type stars down to the apparent median luminosity of the Red
Clump ($K_{s}\approx 15.5$) -- and therefore mostly old and lower mass
stars -- rises much more slowly and may even be flat within a few
$0.1$\,pc of the black hole \citep{SchoedelEtAl07,DoEtAl09}.

Typically, the projected stellar surface number density (and
equivalently the 3D density) is described by a power law of the form
$\Sigma(R)\propto R^{\Gamma}$, where $R$ is the distance from Sgr\,A*.
\citet{BuchholzEtAl09} for example found that the surface density
distribution within a projected radius $R=6"$ of Sgr\,A* can be described with an exponent $\Gamma= 0.2\pm0.1$, which
would correspond to a decreasing density within the innermost few
arcseconds (one arcscond corresponds to approximately 0.04\,pc at the
distance of the GC).

Since the relaxation time at the GC is of order of a Hubble time, old
stars should have had enough time to settle into an equilibrium
distribution around the black hole.  \citet{BW76} found that a
single-mass stellar population should settle into a power-law cusp
$\rho(r) \sim r^{-\gamma}$ with a slope of $\gamma=1.75$ within the influence radius of the
central supermassive black hole, much steeper than what is
observed. This has led to the so-called 'missing giant star'
problem. Moreover, the impact goes beyond stellar dynamics and the
GC because these massive stars close to the MBH are the
precursors of compact objects, which might eventually become a source
of gravitational waves for a space-borne observatory such as LISA or Taiji
\citep{Amaro-SeoaneEtAl2013,Amaro-SeoaneEtAl2012b,GongEtAl2015}. The gradual inspiral of a stellar-mass black hole has been
coined an ``extreme-mass ratio inspiral'', and it accumulates hundreds
of thousands of cycles in the detector band, with impressive
implications in fundamental physics and astrophysics
\citep{Amaro-SeoaneEtAl07,Amaro-SeoaneLRR2012,Amaro-SeoaneGairPoundHughesSopuerta2015}.

A number of works have tried to explain this conundrum with different ideas,
which follow one of three possibilities. When \cite{Genzel1996} first
discovered missing RGB stars, they proposed that this might be due to stellar
collisions depleting giant stars in the innermost parts through the high stellar
densities that are reached near the SMBH. Later, \citet{dav98}, \citet{ale99},
\citet{bai99}, and \citet{dale09} addressed this idea in detail and came to the
conclusion that it can only explain the absence of the brightest and most
extended giant stars. A different suggested possibility to explain the missing
stars is that our GC does not only have one, but a binary of two massive black
holes.  This hypothesised binary could indeed carve a core into the stellar
distribution through three-body interactions, as shown by a number of authors
\citep{bau06,PZ2006,mat07,loc08,gua12}. Nonetheless, the mass of the secondary
needs to be of the order of $\sim 10^5~M_{\odot}$ to explain the observed core.
Such a massive secondary black hole would require the Milky Way to have experienced a major merger relatively recently, which is excluded by observations
\citep[see][]{hansen03,yu03,chen13}. Moreover, the existence of such a massive
secondary black hole is largely ruled out from a number of other
considerations, for instance, constraints on the proper motion of Sgr\,A* from radio
interferometry \citep[see][]{gualandris2009}.
A number of inspiraling smaller-mass black holes can also create a shallow
stellar density profile in the centre, which would relax the major merger requirement, as has been
shown by \citet{mb2014}. It has also been put forward that a star
cluster falling towards the GC could increase the density profile outside of
$10\arcsec$, so that within this distance the profile would be like a core
\citep{kim03,ern09,ant12,antonini2014}. However, mass segregation would rebuild a steeper
profile in as fast as a quarter of the relaxation time \citep[as shown
by][]{PAS2010,ASP2011}.  This requirement would hence need a steady inflow of
clusters to maintain a weak cusp profile in the centre. Finally,
\citet{Merritt2010} and \citet{antonini2014} found that if the nuclear cluster in the GC formed with an extended enough initial core profile, the current stellar
distribution would still not be dynamically relaxed. While this solution is possible,
it requires fine-tuning in the initial conditions to produce the density
distribution seen in the GC. \cite{Amaro-SeoaneChen2014} proposed
that the discs of young stars observed at the GC \citep{paumardetal2006}
are connected to the missing bright giants: the precursor gaseous disc must have gone through a
fragmentation phase that produced dense enough clumps to ensure an efficient
removal of the outer layers of the giants through collisions, rendering them
invisible to observations. Their degenerate cores would nonetheless populate
the same area of phase space where the missing bright giants should be.
\citet{KiefferBogdanovic2016} recently showed that
in order to be viable, this scenario requires the total mass of the fragmenting
disc to have been several orders of magnitude higher than that of the
early-type stars in the stellar discs in the GC.

Owing to the extreme extinction and source crowding towards the GC,
observational studies are complex, even with the power of AO
assisted
10m class telescopes. In particular, the spectroscopic identification
of stars is limited to bright giants and massive young stars, while
the crowding makes it almost impossible to detect main-sequence stars
lower than two solar masses. \citet{Gallegocanoetal2017} and
\citet{Schoedeletal2017} have revisited the observational
data with improved methods. On the one hand, they were able to push the
completeness of the star counts one magnitude deeper than in previous
studies. This means that not only bright giants down to the Red Clump
(RC), but also fainter giants and sub-giants are included in the star
counts. Thus we now have access to two stellar tracer populations,
with different luminosities but similar masses, that can provide us
with information on the structure of the nuclear star cluster
(NSC). Furthermore, the authors also succeeded in deriving the surface
brightness profile of the diffuse stellar light, which traces even
fainter stars, probably sub-giants and main-sequence stars of
$\lesssim1.5\,$M\,$_{\odot}$. The two new analyses provide fully
consistent results. The spatial stellar density can be described by
a power law with exponents between $\gamma 1.15$ to $1.40$ inside 0.5 pc for three
different tracers: resolved giant and sub-giant stars in the magnitude ranges
$12.5\leq K_{s}\leq16.0$ and $17.5\leq K_{s}\leq18.5$ and the diffuse
light, which is created mainly by sub-giants and main-sequence
stars. After combining the data on the central 1-2\,pc with other
measurements of the stellar distribution on scales 2-10\,pc, the
authors de-projected the surface density and determined a 3D power-law
index $\gamma$ between $1.15$ to $1.40$. For RC stars and brighter giants, \citet{Schoedeletal2017} reproduced the findings of previous work, that is, a flattening of the stellar density distribution in the innermost 0.1 to 0.3 pc. The authors concluded that outside of this projected radius, the distribution of these stars is consistent with a stellar cusp, but that some process has probably altered the giant star distributions at smaller radii.

The masses of all tracer populations are very similar, between
$1-2\,$\,M$_{\odot}$. These stars can live for several Gyr and thus be old enough to be dynamically relaxed.
If the faint stars and diffuse stellar light at the GC indeed trace an underlying old stellar population (of age $\sim10^9$ to $\sim10^{10}$ years), then the stellar cusp is surprisingly shallow.  Here it becomes important to look at the theory in more detail. The steep
power-law cusp solution found by \citet{BW76} only holds for a
single-mass stellar population. If stars follow a distribution of masses, mass segregation causes stars of different masses to follow different density distributions with more massive components that have steeper central slopes \citep{BW77,BaumgardtEtAl04b,ASEtAl04,FAK06a}.
In particular, \citet{AlexanderHopman09} and \citet{PAS2010}
independently derived the so-called ``strong-mass segregation''
solution for two stellar mass groups, which is more efficient than the solution expected from the theory of \cite{BW77}, which is mathematically correct but physically inappropriate because of the number fractions
they used.

\citet{Baumgardtetal2005} showed
that in globular clusters containing intermediate-mass black holes,
the surface density distribution of luminous stars can be described by
a weak power-law cusp distribution, offering a new way to explain the
missing-mass problem seen in the GC. However, their
results cannot be directly applied to it, since in
their simulations all stars formed at the same time, which is a valid assumption
for a globular cluster, but not for the nuclear cluster at the centre
of the Milky Way, where stars have formed continuously over a Hubble
time \citep{Pfuhletal2011}. In addition, while the GC has
a relaxation time of the same order as its age, the star clusters in their simulations were simulated for about ten relaxation times.

In this paper we present results of direct $N$-body simulations aimed at studying the dynamics
of the nuclear cluster in the Milky Way. In our simulations, we evolve a star cluster surrounding
a central massive black hole over a Hubble time under the combined influence of two-body relaxation and continuous star formation. Our paper is organised
as follows: In section 2 we discuss our simulations and in section 3 we compare the resulting distribution
of stars with observations of the nuclear star cluster in the Galactic centre.
Section~4 presents our conclusions.

\section{Simulations}

We have run $N-$body simulations of star clusters containing massive central
black holes using the GPU enabled version of the collisional $N$-body code {\it NBODY6} \citep{Aarseth99,NitadoriAarseth2012},
and we followed the evolution of the star clusters under the combined influence of star formation, stellar evolution, and two-body relaxation.
The mass distribution of stars in our simulations was given by a \citet{Kroupa01} mass function with lower and upper mass limits of
0.1 and 100 M$_\odot$ , respectively. This mass function is compatible with observations of the mass-to-light ratio of the nuclear cluster and the temperature and luminosity distribution of
individual giant stars in the nuclear cluster \citep{loeckmannetal2010,PfuhlEtAl2015}.
Owing to the high escape velocity from the nuclear cluster, we assumed a 100\% retention fraction for stellar-mass black holes and neutron stars upon their formation. We also assumed solar metallicity for the stars in our simulation, which is compatible with the average metallicity of stars in the GC \citep{rs2015,doetal2015}.

Because of the high computational cost of $N$-body simulations, we simulated clusters with smaller particle numbers but larger half-mass radius so that the relaxation time of the
simulated clusters is the same as the relaxation time of the nuclear cluster in the GC. This approach is similar to the approach used by \citet{Baumgardtetal2003} to
model the globular cluster G1 and \citet{Baumgardt2017} to model Galactic globular clusters. It allows us to correctly model the effects of two-body relaxation, which is the
main focus in this paper, but processes that do not occur on relaxation timescales like tidal disruption of stars cannot be modelled easily in scaled simulations.
The nuclear cluster in the GC has a projected half-light radius of $R_{h NC} = 4.2 \pm 0.4$ pc and a mass of $M_{NC} = 2.5 \pm 0.4 \cdot 10^7$ M$_\odot$ \citep{SchoedelEtAl2014}. Using the
definition of the half-mass relaxation time as given by \citet{Spitzer87} and assuming the 2D half-light radius to be 75\% of the 3D half-light
radius, this results in a half-light relaxation time of
$T_{RH} \approx 14$ Gyr. We therefore used an initial projected half-mass radius of our simulated clusters
of $R_h = R_{h NC} \cdot (M_{NC}/M_C)^{1/3}$ = 26.9 pc, so that our simulated clusters have the same half-mass relaxation time as the nuclear cluster and evolve dynamically with the same rate per physical time. At the end of the simulations, we scaled our clusters down to a projected half-light radius of $R_{h NC}=4.2$ pc, so that we were able to directly compare with observational data. In total we performed three realisations using different random number seeds
for each density profile and overlaid the results of five snapshots of each realisation centred on the MBH
after scaling each cluster to the same half-mass radius.

\citet{Pfuhletal2011} found that most stars in the nuclear cluster are old, with about 80\% of them having been born more than 5 Gyr ago. They also found that an exponentially decreasing star formation rate according to
\begin{equation}
SFR(t) = e^{-t/\tau_{SFR}}
\label{eq:sfr}
\end{equation}
with a characteristic timescale $\tau_{SFR}$ = 5.5 Gyr matches the age distribution of old stars in the nuclear cluster. We therefore started our simulations with a star cluster
containing $N=50,000$ stars and an MBH that contained 15\% of the cluster mass, similar to the mass ratio seen for the GC SMBH \citep[see e.g.][ and references therein]{GenzelEtAl10}. We then simulated the evolution of this cluster for 1~Gyr and added new stars to it after 1~Gyr with a rate given by Eq.~\ref{eq:sfr}. We evolved the new cluster for another Gyr before adding the next generation of stars and repeated this process until the cluster reached an age of $T=13$ Gyr. Throughout the evolution, the mass ratio between the cluster and the central MBH was kept constant at 15\%, that is, we assumed that the MBH grows at the same rate as the nuclear cluster. The new stars added to the simulation following a \citet{King66} model with dimensionless central potential of $W_0=5.0$ initially.
In order to study the influence of the initial density profile on the final results, we also ran simulations in which the initial density
profiles had central potentials of $W_0=3.0$ and $W_0=7.0$, but found that the final slopes of the power-law profiles do not change by
more than $\Delta \Gamma = \pm 0.1$ in surface density, confirming that the GC nuclear star cluster is dynamically relaxed
in its centre. When setting up the \citet{King66} models, we took the potential of the central black hole into account when calculating
the stellar velocities. After 13 Gyr, we ended up with star clusters containing about $\sim255,000$ stars and a final mass of $M_C=9.5 \cdot 10^4$ M$_\odot$.
\begin{figure}
\resizebox{\hsize}{!}
          {\includegraphics[scale=1,clip]{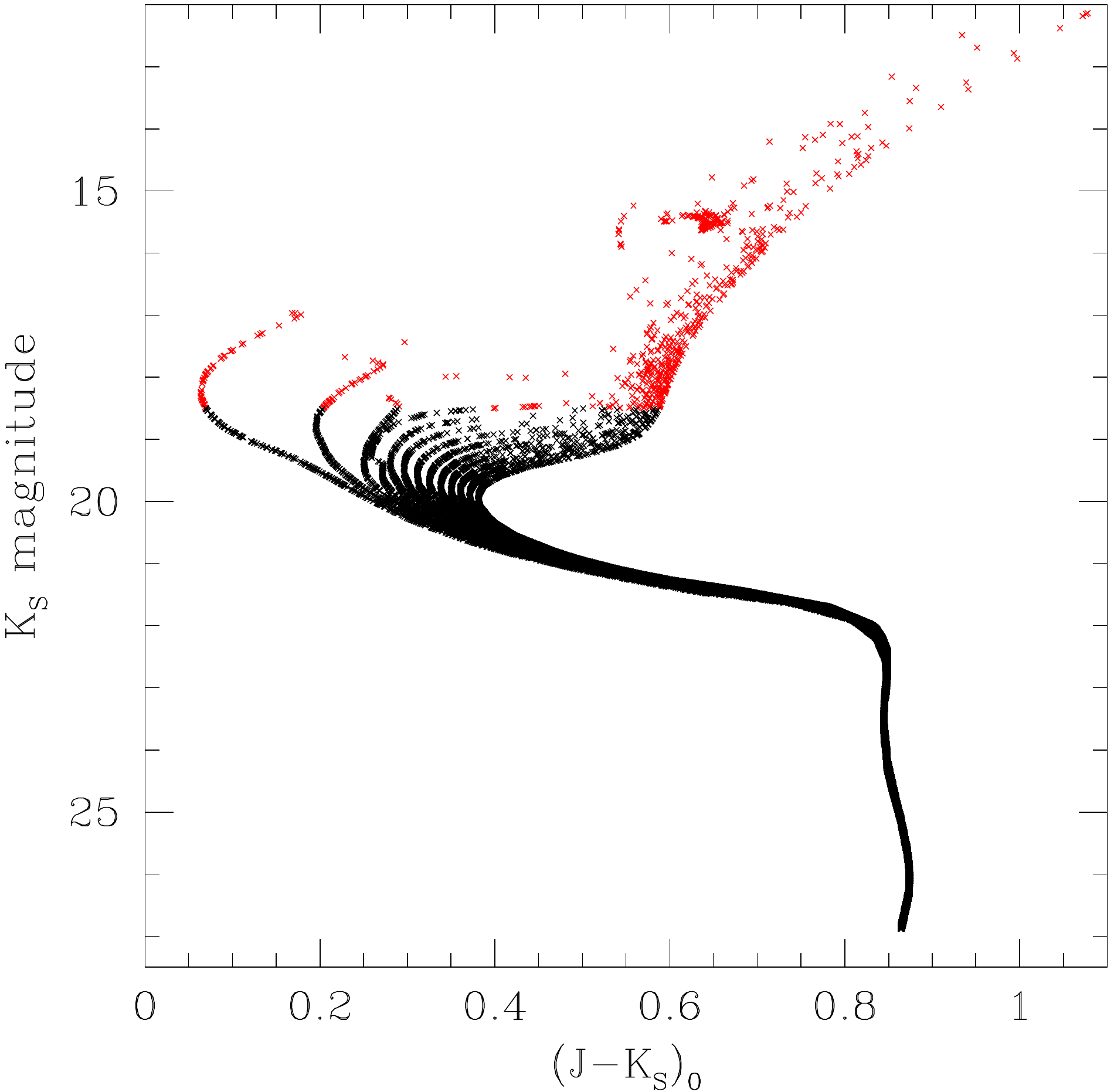}}
\caption
   {
$J-K_S$ CMD of the simulated nuclear clusters after 13 Gyr of evolution for a distance of 8~kpc to the GC and and an average reddening of $A_{Ks}=2.54$ mag. The different stellar generations that were added to the simulations can clearly be distinguished in the CMD. Stars marked in red have $K_S<18.5$ and are used for comparison with the density profile of resolved stars from \citet{Gallegocanoetal2017}. The remaining stars are used to create the diffuse light profile of the simulated clusters.
   }
\label{fig.phot}
\end{figure}

Figure~\ref{fig.phot} depicts a J-$K_S$ vs $K_S$ colour-magnitude diagram (CMD) of the stars in our simulation after $T=13$ Gyr, calculated using the PARSEC isochrones \citep{Bressanetal2012}. In order to calculate the apparent $K_S$ band magnitudes, we assumed a distance of 8 kpc to the GC and a $K_S$ -band extinction of $A_{Ks} = 2.54$ mag \citep{Schoedeletal2010} for all stars. The different generations of stars that were added to the star cluster during the simulation can clearly be distinguished in the CMD. In the following, we use stars with $K_S < 18.5$ mag (shown in red) for the comparison with the number density profile of resolved stars from \citet{Gallegocanoetal2017}. For the comparison with the diffuse light profile of \citet{Schoedeletal2017}, we used all stars fainter than this limit and summed their $K_S$ -band luminosities. With these limits,
resolved stars have masses in the range $0.86 < m < 2.31$ M$_\odot$ , while 90\% of the diffuse light is created by stars with masses $0.78 < m < 1.76$ M$_\odot$. On average, the resolved stars are about 0.25 $M_\odot$ more massive than the brightest stars that contribute to the unresolved light.

\begin{figure}
\resizebox{\hsize}{!}
          {\includegraphics[scale=1,clip]{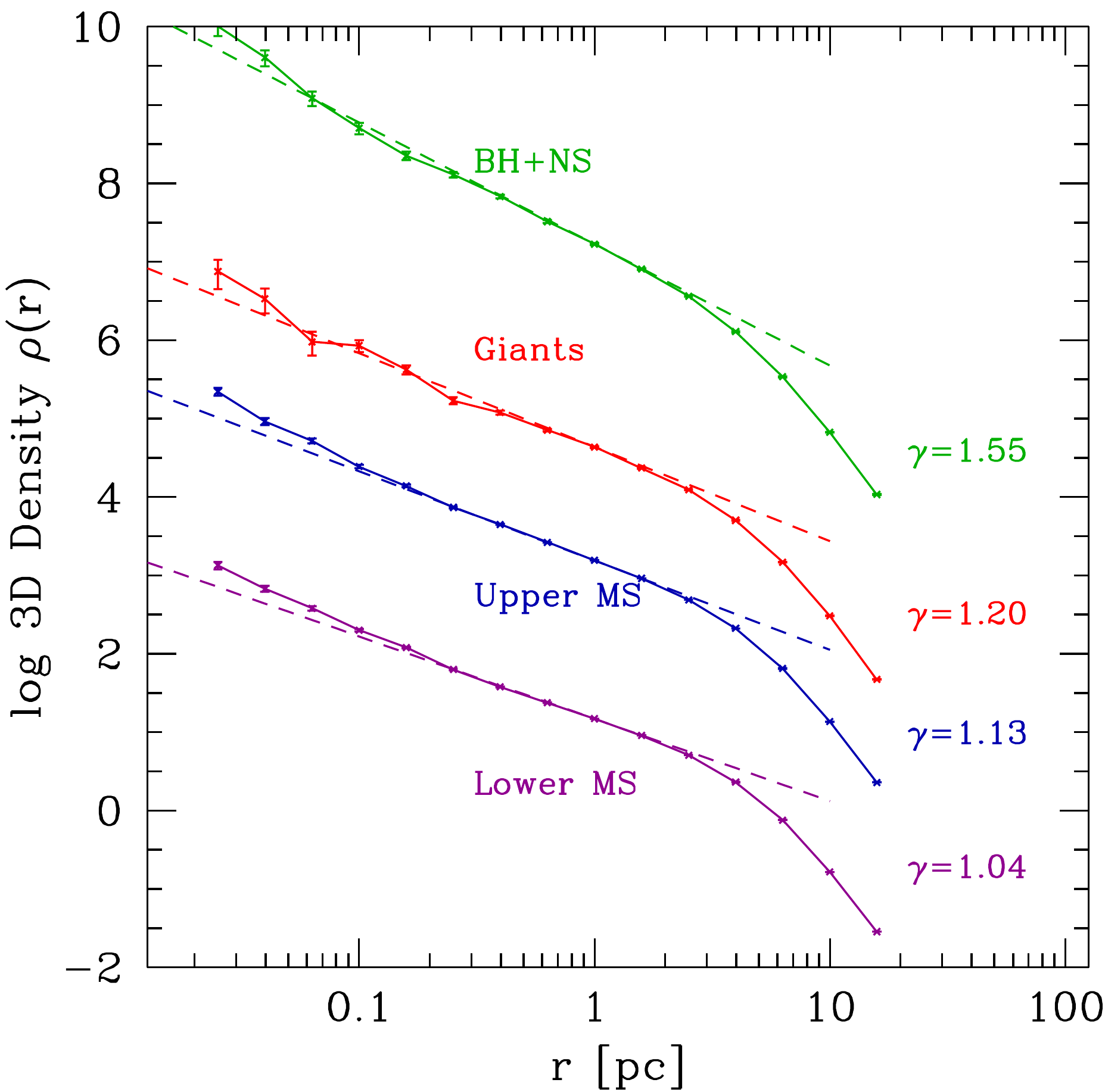}}
\caption
   {
Spatial density of different stellar components at the end of our simulations.
From top to bottom, we show the density distribution of black holes and neutron stars
(green), giant stars with apparent magnitudes $K_S<18.5$ (red), upper main-sequence
stars with masses 0.6 M$_\odot$ and 0.8 M$_\odot$ (blue), and low-mass main-sequence stars
with masses 0.1 and 0.3 M$_\odot$ (orange). For clarity, curves are shifted vertically.
Dashed lines show power-law density distributions $\rho(r) \sim r^{\gamma}$ fitted to the density distribution inside the influence radius of the black hole ($r_{BH}=2.8$ pc). Because of mass segregation, the slope steepens from $\gamma=1.04$ for the lowest-mass main-sequence stars to $\gamma=1.55$ for black holes.
   }
\label{fig.rho3d}
\end{figure}

\section{Results}

Figure~\ref{fig.rho3d} shows the density distribution of different types of
stars after 13 Gyr of evolution. Shown are the density distributions of
stellar mass black holes; giant stars, which we assume to be all stars with
K-band luminosities $K_S<18.5$; upper main-sequence stars with masses
between 0.6 M$_\odot$ and 0.8 M$_\odot$ ; and lower main-sequence stars
with masses between 0.1 and 0.3 M$_\odot$. For clarity we have shifted all
curves vertically to separate the different curves from each other. We also
scaled the final cluster to have a projected half-light radius of 4.2 pc to match
the GC nuclear cluster. For  this half-light radius, the influence radius of the central black hole, that is, the radius where the cumulated mass in
stars becomes equal to the central black hole mass, is $r_{BH}=2.8$
pc. This value is close to the break radius seen in the surface density distributions of resolved stars and diffuse light
as found by \citet{Gallegocanoetal2017} and \citet{Schoedeletal2017} and depicted in Figs.~3 and~4.

\begin{figure}
\resizebox{\hsize}{!}
          {\includegraphics[scale=1,clip]{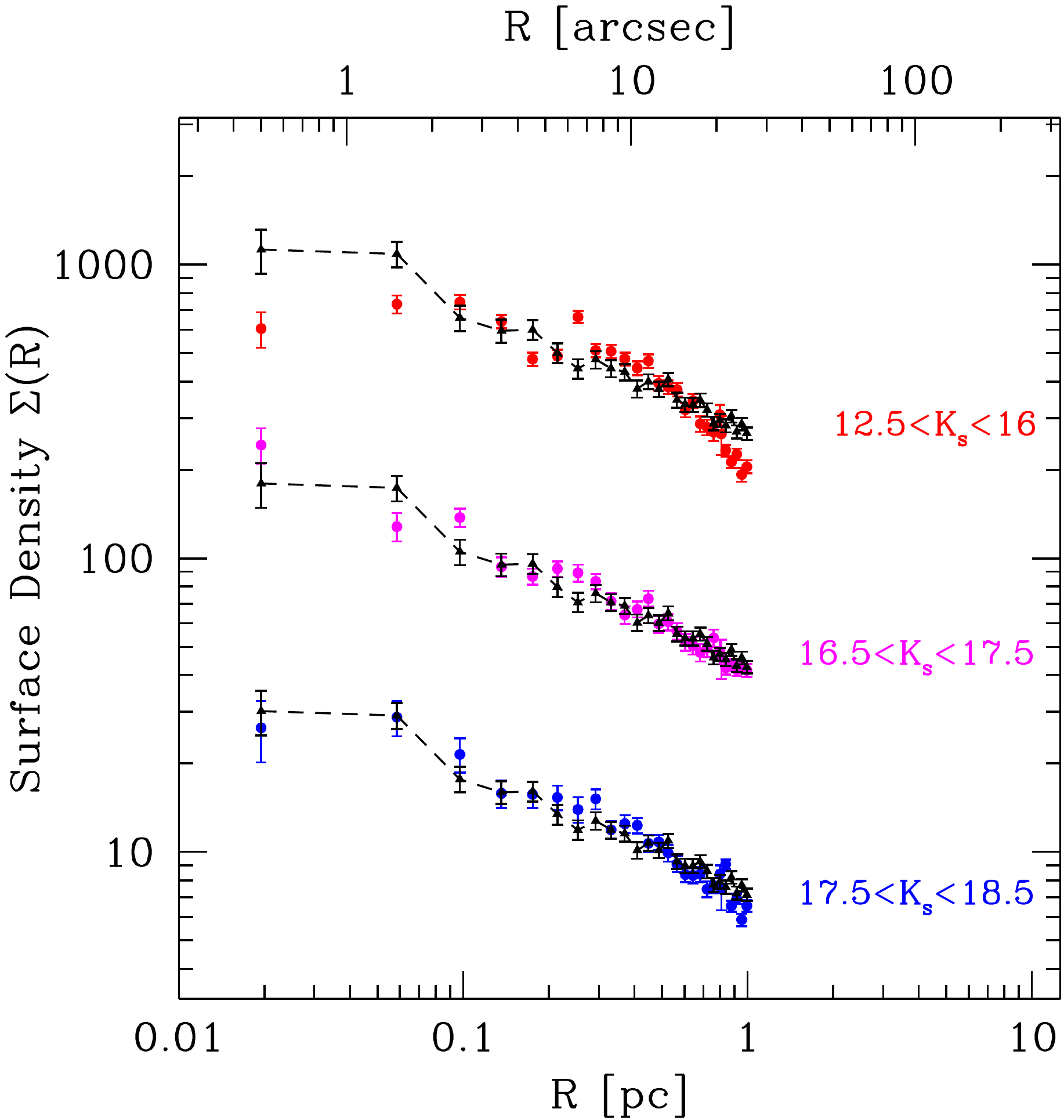}}
\caption
   {
Surface density $\Sigma(R)$ of old giant stars in the GC for three
different $K_S$ -band intervals. The observed surface densities and associated error bars are taken from \citet{Gallegocanoetal2017} and are shown by
circles for the different magnitude intervals. Black dashed lines and triangles show the profile that we obtain for giant stars at the end of our simulations. The profiles are in excellent agreement with each other in the different magnitude intervals, with the exception of the innermost density profile of stars in the magnitude interval $12.5 < K_S < 16$.
   }
\label{fig.gallego_res}
\end{figure}

Inside the influence radius, the density distribution of the
different stellar components can be well described by single-power-law distributions.
Black holes and neutron stars follow a power-law distribution with the steepest slope
of $\gamma=1.55$, close to the theoretically predicted slope of $\gamma=1.75$
for a single-mass stellar population around a massive black hole \citep{BW76}.
Owing to mass segregation, all other components follow flatter density
distributions. The slopes vary by only
$\Delta \gamma = 0.16$ between giant stars and the lowest-mass main-sequence stars, however. Overall, the power-law slopes of the different stellar
mass groups are flatter than what \citet{BaumgardtEtAl04b} found for a
single-age stellar cluster with the same initial mass function after
$T=12$ Gyr of evolution (see their Fig.~7). This indicates that the
GC star cluster is dynamically not yet completely relaxed
because of its relatively long relaxation time ($T \sim 14$ Gyr). In
addition, the continuous star formation reduces mass segregation between
the different stellar components.

Because of
the small amount of mass segregation, low-mass stars dominate the stellar mass profile for all radii outside 0.01 pc and the luminosity profile closely follows the mass profile in the nuclear cluster, that is, we do not predict a strong variation of the $M/L$ profile of the nuclear cluster with radius. When scaled to the GC, our simulations predict about
$\sim 300$ stellar-mass black holes and a total mass of 7000 M$_\odot$ in stars
inside the central 0.1 pc. Our mass estimate is in good agreement with the
approximate $10^{4}$\,M$_{\odot}$ of total mass within this region that can be calculated with the Nuker model and the mass normalisations of \citet{Schoedeletal2017}, and it is also consistent with the upper limit of $1.3 \cdot 10^5$ M$_\odot$ recently derived by \citet{boehleetal2016}. We note that these mass limits have been derived assuming a  mass-to-light ratio that is independent of radius for the nuclear star cluster.

Figure~\ref{fig.gallego_res} depicts the distribution of bright stars in the central parsec and compares it with the observed distribution of late-type
giant stars within three different $K_S$-band intervals as determined by \citet{Gallegocanoetal2017}. In order to avoid uncertainties and systematic biases in mixing different observational data sets, we only depict the NACO data, that is, data in the radial range from 0.011~pc to 1~pc in Fig.~\ref{fig.gallego_res}. In order to increase the statistical significance of our results, we use all stars with $K_S<18.5$ in the simulations
for the comparison with the observed distribution in the three magnitude intervals. This is justified since the observed
stars are giant stars, which show only a small dependence of average mass and age with luminosity. Hence we
expect the stars in the different magnitude bands to follow similar density distributions in our simulations.
The observed and simulated density profiles agree very well with each other, the differences between the two profiles
are usually less than 10\% for all three magnitude intervals and are of the same order as the uncertainties of the observed profiles.
The combined surface density distribution of the three simulations we have performed can be fitted by a power-law distribution $\Sigma(R) \sim R^{-\Gamma}$ with a slope $\Gamma \sim 0.46$
in the range 0.04 pc $< R <$ 1.0 pc. This is in excellent agreement with the observed density distributions of bright stars, which have best-fitting slopes of $\Gamma = 0.45 \pm 0.01$
for stars with $12.5 < K_S < 16$, $\Gamma = 0.47$ for stars with $16.5 < K_S < 17.5$ and $\Gamma = 0.47 \pm 0.02$ for stars with $17.5 < K_S < 18.5$ over the same radial range \citep{Gallegocanoetal2017}.
The observed data have almost the same slopes in the different bands, as expected from our simulations. While we compared our results only to the most recent results, we
note that previous determinations of the surface density distribution of late-type giant stars \citep[e.g.][]{SchoedelEtAl07,BuchholzEtAl09} are also compatible with
our results, at least outside the central few arcsec.
Stars with $12.5 < Ks < 16$ display a density deficit inside the innermost few arcsec, while their distribution at larger radii agrees with our simulations. This could indicate that some depletion process like giant star collisions or disc passages has altered their apparent distribution at the very centre.

\begin{figure}
\resizebox{\hsize}{!}
          {\includegraphics[scale=1,clip]{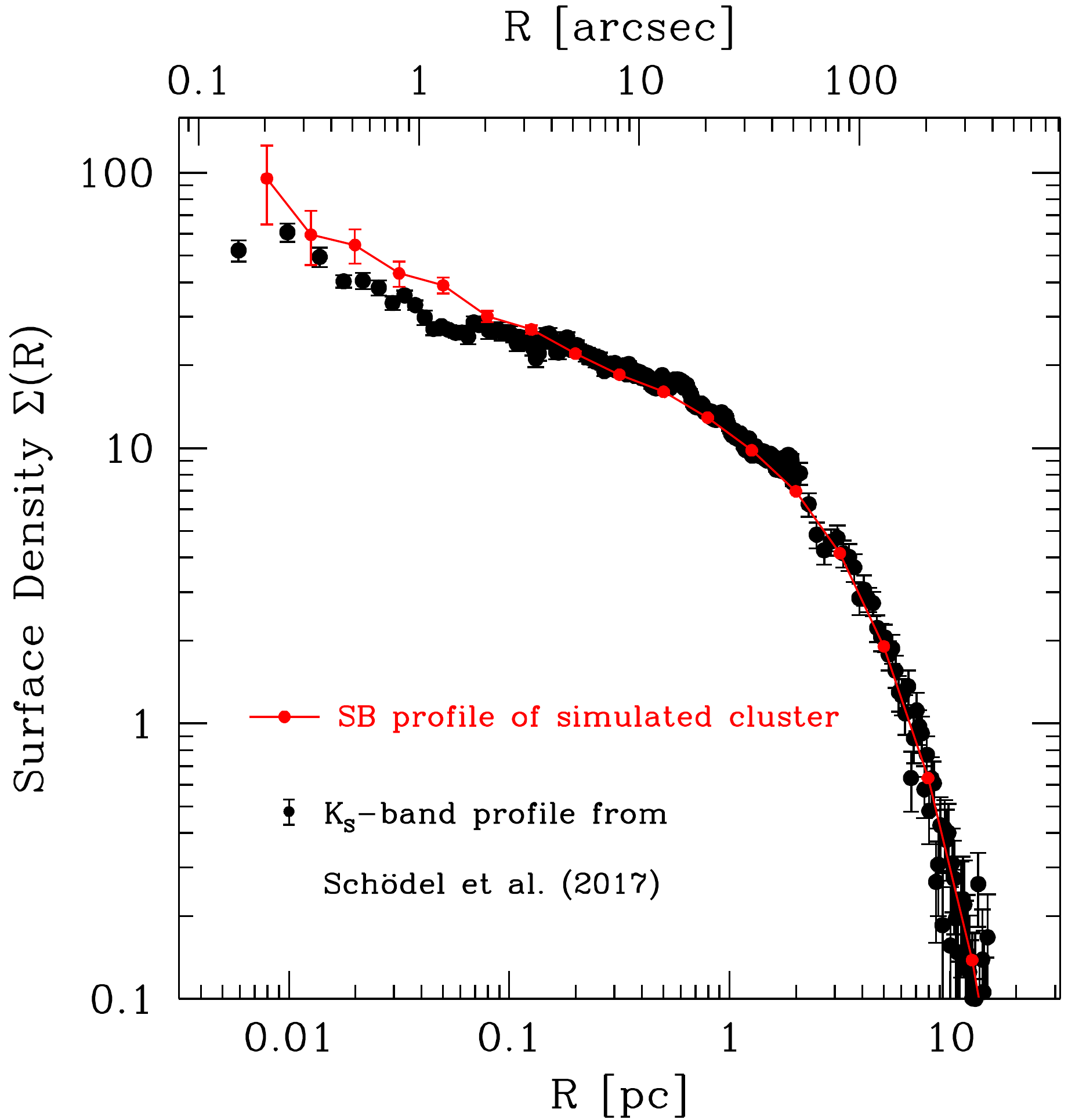}}
\caption
   {
$K_S$ -band surface luminosity profile $\Sigma(R)$ of the diffuse light of the GC star cluster. The data points show the results of \citet{Schoedeletal2017} corrected for differential extinction. Red line and data points show the surface luminosity profile of the simulated cluster, determined from all main-sequence stars fainter than $K_S>18.5$. Theoretical and observational profiles are in excellent agreement with each other outside the innermost few arcsec.
   }
\label{fig.schoedel_diff}
\end{figure}
Figure~\ref{fig.schoedel_diff} compares the surface density profile of
diffuse light in the GC from \citet{Schoedeletal2017}
with the results of our simulations. The black data points depict the
$K_S$ -band surface luminosity derived by \citet{Schoedeletal2017}
after correction for differential extinction and combining their data inside 1.5 pc with the
azimuthally averaged extinction-corrected near-infrared data by \citet{fritzetal2016}. The red line shows the
surface luminosity density of faint stars in our simulations. The
simulated surface light distribution is well described by a single power-law profile with
$\Sigma(R) \sim R^{-0.37}$ inside 1.0 pc. This is close to the observed profile,
which has $\Gamma = 0.26 \pm 0.02_{stat} \pm 0.05_{sys}$ \citep{Schoedeletal2017}. The slope predicted from our simulations
is nearly the same as in the case of the resolved stars, again indicating that there was
insufficient time for the nuclear cluster to develop significant mass
segregation between main-sequence stars of different masses. The observed profile is slightly below our
predicted data in the innermost few arcsec, but the differences are within the uncertainty with which we can determine the
surface luminosity profile from our simulations
and might therefore not be significant. In addition, the surface density of the diffuse light may show some systematic uncertainties at radii R = 1-2''
because of the presence of very bright stars and the related difficulty of accurate PSF subtraction in this region (see discussion in \citealt{Schoedeletal2017}).
We also see that our model somewhat over-predicts the surface density of diffuse light at radii $R < 2$''. However, the qualitative agreement, in particular the
slope of the power-law at larger radii, is very good. There is a clear steepening of both the observed and simulated profiles beyond 20 arcsec, that is,
outside the influence radius of the central black hole.

We conclude that the simulated and observed data agree excellently well with each other. In particular, the inferred power-law indices of the surface densities as well as the corresponding 3D power-law slopes (see caption of Fig. 2) agree well with the values estimated by \citet{Gallegocanoetal2017} and \citet{Schoedeletal2017}.
The surface density profile of both resolved stars and diffuse light of the nuclear cluster are therefore entirely compatible with what we expect for a star cluster around a massive black hole that is evolving under the combined influence of dynamical relaxation and continuous star formation given the age and star formation history of the nuclear cluster.

\section*{Conclusions}

The existence of stellar cusps in relaxed clusters around massive black
holes is a long-standing prediction of theoretical stellar
dynamics, but it has escaped confirmation for decades for several
reasons, some observational and some theoretical. On the observational
side, there is only a single target where the cusp theory can be
tested with current instrumentation: the Galactic
centre. Extragalactic nuclei are too far away for current
instrumentation, and the existence of (intermediate-mass) massive black
holes in globular clusters is still debated. However, the GC is not an
easy target, and number counts and stellar classification both suffer
from the very high and spatially highly variable interstellar
extinction as well as from the extreme source crowding.
Observational
evidence has therefore been ambiguous for a long time. Recent new
analyses, however, have found a stellar cusp around Sgr\,A*, with very
consistent morphologies for different tracer populations that
differ by several magnitudes in brightness. While the distribution of
resolved giant stars with $K_S=18$ might be contaminated by young stars that formed
within the last few 100 Myr, the contribution of such stars is much smaller for
brighter giant stars with $K_S<16$ and the diffuse stellar light \citep{Gallegocanoetal2017}.
The observations therefore imply the existence of a stellar cusp among the
old and dynamically relaxed giant stars.
This cusp has been found to be rather shallow, which is one of the
reasons why it has so successfully escaped detection over a long time.

On the theoretical side, the main limitation was probably that the
systems that were analysed were too simple. As we now know, the nuclear cluster around the SMBH at
the GC is a highly complex system and has experienced many episodes of repeated
star formation and/or cluster infall. As a result of the different physical and dynamical
ages, stars that formed at different epochs will follow different radial distributions
even if their masses are the same \citep{AharonPerets15}. This work presents an attempt to
take the complex star formation history of the Milky Way's NSC into account. As
we have shown, two-body relaxation and repeated star formation across the lifetime of the Galaxy indeed results in a rather weak stellar cusp, consistent with the observations.
It appears that this is the first time that theory and observations reach
convergence on the question of the stellar cusp. Because we can be sure
now that there is a power-law cusp around Sagittarius A*, the
apparent deficit of giants within the innermost few arcsec of the central black
hole also indicates that the latter do not trace the overall cluster structure.
Either they do not have the same age structure as the fainter stars, or some
process - such as a destruction of their envelopes via interactions with a fragmenting gaseous disc
\citep[the idea of][]{Amaro-SeoaneChen2014} - has altered their apparent distribution.

The existence of a stellar cusp at the heart of the Milky Way implies
that density cusps might also exist in many other galactic nuclei, especially
smaller nuclei with relaxation times shorter than a Hubble time. This
is of importance for gravitational wave astronomy because
it means that Extreme-Mass Ratio Inspirals (EMRIs) may be observed with significant
frequency.


\begin{acknowledgements}
The research leading to these results has received funding from the European
Research Council under the European Union's Seventh Framework Programme
(FP7/2007-2013) / ERC grant agreement $n^{\circ}$ [614922].
PAS acknowledges support from the Ram{\'o}n y Cajal Programme of the Ministry
of Economy, Industry and Competitiveness of Spain. This work has been partially
supported by the CAS President's International Fellowship Initiative.
\end{acknowledgements}


\label{lastpage}
\end{document}